\documentclass[11pt]{article}
\usepackage{graphicx}
\usepackage{amssymb,amsmath}

\newcommand{\BABARPubYear}    {06}

\newcommand{\BABARConfNumber} {013}
\newcommand{\SLACPubNumber} {11978}

\def\dstarelnubulo{\ensuremath{\Dstarm\ell^+\nu_{\ell}}\xspace}

\def\tag{\ensuremath{\mathrm{tag}}\xspace}
\def\rec{\ensuremath{\mathrm{rec}}\xspace}
\def\qoverp{\ensuremath{\bigg\vert\frac{q}{p}\bigg\vert}\xspace}
\def\qoverp2{\ensuremath{\bigg\vert\frac{q}{p}\bigg\vert^2}\xspace}
\def\poverq2{\ensuremath{\bigg\vert\frac{p}{q}\bigg\vert^2}\xspace}
\def\mnu{\ensuremath{M_\nu^2}\xspace}

\def\damt{\ensuremath{\cdot 10^{-3}}\xspace}
\def\damq{\ensuremath{\cdot 10^{-4}}\xspace}
\def\psoft{\ensuremath{\pi_s}\xspace}

\input babarsym
 
\long\def\inst#1{\par\nobreak\kern 4pt\nobreak
    {\it #1}\par\vskip 10pt plus 3pt minus 3pt}

\begin{document}
{\pagestyle{empty}

\begin{flushright}
\babar-CONF-\BABARPubYear/\BABARConfNumber \\
SLAC-PUB-\SLACPubNumber \\
July 2006 \\
\end{flushright}

\par\vskip 4cm

% Title of the paper
\begin{center}
\Large \bf A measurement of  \boldmath{$CP$}-violation parameters in \boldmath{$\BzBzb$} mixing 
using partially reconstructed \boldmath{\dstarelnubulo} events at \babar\ 
\end{center}
\bigskip

\begin{center}
\large The \babar\ Collaboration\\
\mbox{ }\\
\today
\end{center}
\bigskip \bigskip

% Abstract
\begin{center}
\large \bf Abstract
\end{center}
$CP$ violation in \BzBzb\ mixing is characterized by the value of the parameter
$|q/p|$ being different from 1, and the Standard Model predicts this 
difference to be smaller than $10^{-3}$. We present a measurement of this
parameter using a partial reconstruction of one of the $B$ mesons in the 
semileptonic channel \dstarelnubulo, where only the hard lepton and the soft 
pion from the $\Dstarm \rightarrow \Dzb \pi^-$ decay are reconstructed. 
The flavor of the other $B$ is determined by means of lepton tagging. 
The determination of $|q/p|$ is then performed with a fit to the proper time
difference of the two $B$ decays.
We use a luminosity of 200.8 $\mathrm{fb}^{-1}$, collected 
at the \FourS\ resonance by the \babar\ detector at the \pep2\ 
asymmetrical-energy \epem\ collider, in the period 1999-2004. We obtain 
the preliminary result:
\begin{equation}
\nonumber
|q/p| - 1 = (6.5 \pm 3.4(\mathrm{stat.}) \pm 2.0(\mathrm{syst.})) \damt 
\end{equation}
\vfill
\begin{center}

Submitted to the 33$^{\rm rd}$ International Conference on High-Energy Physics,
ICHEP 06,\\
26 July---2 August 2006, Moscow, Russia.

\end{center}

\vspace{1.0cm}
\begin{center}
{\em Stanford Linear Accelerator Center, Stanford University, 
Stanford, CA 94309} \\ \vspace{0.1cm}\hrule\vspace{0.1cm}
Work supported in part by Department of Energy contract DE-AC03-76SF00515.
\end{center}

\newpage
} 

% Input author list file
\begin{center}
\small

The \babar\ Collaboration,
\bigskip

%% author list as of 01-Jul-2006 (596 authors)
%
{B.~Aubert,}
{R.~Barate,}
{M.~Bona,}
{D.~Boutigny,}
{F.~Couderc,}
{Y.~Karyotakis,}
{J.~P.~Lees,}
{V.~Poireau,}
{V.~Tisserand,}
{A.~Zghiche}
\inst{Laboratoire de Physique des Particules, IN2P3/CNRS et Universit\'e de Savoie,
 F-74941 Annecy-Le-Vieux, France }
% ...

\end{center}\newpage

\section{INTRODUCTION}
\label{par:theo}
Although $CP$ violation has been established both
in the interference between decay and mixing and in direct $B$ decay, $CP$ 
violation in the mixing alone has up to now eluded experimental observation.

In the standard mixing formalism, the effective Hamiltonian is expressed
as the sum of a mass and a decay matrix ($\mathbf{H} = \mathbf{M}
- i/2~\mathbf{\Gamma}$) and the $B$ mass eigenstates are connected to the
flavor eigenstates by:
\begin{eqnarray}
\nonumber
&& |B_L \rangle = p |\Bz \rangle + q |\Bzb \rangle    \\
&& |B_H \rangle = p |\Bz \rangle - q |\Bzb \rangle.
\end{eqnarray}

The absolute value of the ratio $q/p$ can be written in this notation as:
\begin{equation}
\qoverp2 = \Bigg\vert\sqrt{\frac{M_{12}^*-i/2~\Gamma_{12}^*}{M_{12}-i/2~
\Gamma_{12}}}\Bigg\vert^2 \simeq 1 - \mathrm{Im}\left(\frac{\Gamma_{12}}
{M_{12}}\right)
\end{equation} 
so it is exactly equal to 1 in a $CP$-conserving scenario (where the 
off-diagonal elements of the mass and decay matrices are real, that is 
$M_{12} = M_{12}^*$, $\Gamma_{12} = \Gamma_{12}^*$), while it differs 
by a small quantity if $CP$ is violated in mixing. The current theoretical 
predictions on this quantity 
in the Standard Model (SM) are $2 \damq \lesssim |q/p| -1 \lesssim 6 \damq$ 
\cite{ciuch,bene}.

However, recent theoretical publications \cite{laplace} have pointed out 
that in some 
New Physics (NP) general scenarios, the predictions for this quantity can 
be significantly different with respect to the SM. 
Making only the assumptions that the CKM \cite{ckm} is a $3 \times 3$ 
unitary matrix, and the tree-level processes are dominated by the SM,
the matrix element $M_{12}$ in this scenario can be 
related to the SM one by the formula:
\begin{equation}
M_{12}^{\mathrm{NP}} = r_d^2 e^{2i \theta_d} M_{12}^{\mathrm{SM}}
\end{equation} 
where $r_d$ and $\theta_d$ are general New Physics amplitude and phase, while 
$\Gamma_{12}$ is not modified. The $CP$ asymmetry in mixing can be written as:
\begin{eqnarray}
\nonumber
&& A_{SL} = \frac{\Gamma(\Bzb \rightarrow \ell^+ X) - \Gamma(\Bz \rightarrow 
\ell^- X)}{ \Gamma(\Bzb \rightarrow \ell^+ X) + \Gamma(\Bz \rightarrow \ell^- 
X)} \simeq \\ 
&& \simeq 2(1-|q/p|) = -\mathrm{Re}\left(\frac{\Gamma_{12}}{M_{12}}\right)^
{\mathrm{SM}} \frac{\sin{2 \theta_d}}{r_d^2} + \mathrm{Im}\left(\frac
{\Gamma_{12}}{M_{12}}\right)^{\mathrm{SM}} \frac{\cos{2 \theta_d}}{r_d^2} 
\label{eq:np}
\end{eqnarray} 

If the New Physics phase is significantly different from 0, the real 
part of  $\Gamma_{12}/M_{12}$ can become dominant, enhancing the asymmetry 
up to an order of magnitude. Together with the measurements
of CKM parameters, an accurate determination of $|q/p|$ is therefore
an additional constraint on New Physics models.

Measurements of $|q/p|$ at the $B$-factories are performed both
using inclusive dilepton events \cite{belle,yeche} or $B$ mesons fully 
reconstructed into flavor or $CP$ eigenstates \cite{fmv}. The most precise 
result obtained up to now from the inclusive dilepton method 
is $|q/p| -1 = (-0.8 \pm 2.7 \pm 1.9) \damt$ \cite{yeche},
where the first uncertainty is statistical, the second systematic. 
Analogous techniques have also been used by the D$\emptyset$ experiment at the 
Tevatron \cite{dzero}: the
measured parameter in this analysis is not a determination of $|q/p|$ alone, 
but also involves contributions from the $CP$ violation in $B_s$ 
mixing \cite{utfit}, and this has allowed to set constraints also on
New Physics in the $B_s$ sector.  

In this analysis we exploit the partial reconstruction of 
$\Bz \rightarrow \dstarelnubulo$ events. 
Though the total statistics is not 
as high as in the dilepton case, we can keep the charged $B$ background  
at a lower level, while selecting a greater number of events 
than in an analysis requiring the full reconstruction of a 
hadronic or semileptonic decay as a tag for the $B$ flavor. At the same time, 
since the 
reconstructed and tag side are well defined, a procedure to determine 
particle detection asymmetries from data (explained in
Sec.~\ref{sec:TDanalysis}) can be carried out. In this way, we do not need
dedicated control samples to determine the asymmetry induced by the 
experimental cuts.

\section{THE \babar\ DETECTOR AND DATASET}
\label{sec:babar}
The data used in this analysis were collected with the \babar\ detector
at the \pep2\ asymmetrical-energy \epem\ storage ring in the 
period 1999-2004; they correspond
to an integrated luminosity of 200.8 $\mathrm{fb}^{-1}$ (i.e. about 110 
million \BzBzb\ pairs) taken at the mass of the \FourS\ resonance
($\sqrt{s}$ = 10.58 GeV), plus 21.6 $\mathrm{fb}^{-1}$ taken about 
40 MeV below. Samples of simulated $\FourS \rightarrow \BzBzb$ and \BpBm\ 
events are used to estimate efficiencies, study 
background and detector asymmetries. 

The \babar\ detector is described in detail elsewhere~\cite{ref:babar}. 
Tracking of charged particles is provided by a five-layer
silicon vertex tracker (SVT) and a 40-layer drift chamber (DCH). 
Vertices and soft pion tracks are mainly reconstructed using information
from the SVT. Cherenkov radiation detected in 
a ring-imaging detector (DIRC) is used for particle identification. An 
electromagnetic calorimeter (EMC), which consists of 6580 thallium-doped CsI 
crystals, is used to measure electron energies.
These systems are mounted inside a 1.5 T solenoidal
superconducting magnet. The flux return of the magnet (IFR) is equipped with 
Resistive Plate Chambers providing muon identification. 

\section{SELECTION METHOD AND SAMPLE COMPOSITION}
\label{sec:Analysis}

The decay rates for neutral $B$ mesons can be calculated theoretically
as a function of $\Delta t$, the time difference between the decays
of the two \Bz\ mesons, 
taking into account both the time evolution of the mass eigenstates and 
the fact that they are produced coherently from the \FourS\ resonance.
We approximate $\Delta t \sim \Delta z / (\langle\beta\gamma\rangle c)$, 
where $\Delta z$ is 
the measured distance between the decay vertices 
projected along the beam direction ($z$ axis) and 
$\langle\beta\gamma\rangle$
is the average boost of the \FourS\ in the laboratory frame.
The method used to determine $|q/p|$ is a two-dimensional fit to the 
set of variables ($\Delta t$, $\sigma_{\Delta t}$) with a binned 
extended maximum likelihood approach, where  $\sigma_{\Delta t}$ is the 
per-event error calculated on $\Delta t$.

To discriminate events with a \BB\ pair from events which originate from 
the production of
light quarks, we require the ratio of the second to the zeroth Fox-Wolfram
moment \cite{foxwol} $R_2 < 0.5$. The total number of charged tracks 
in the event is required to be greater than 4, to discard lepton pair
production. We also calculate the invariant mass 
of the two highest-momentum leptons in the event and we apply a veto on the 
regions $M_{ll} < 0.35$ GeV/$c^2$ (converted photon rejection) and 
$3.07 < M_{ll} < 3.14$ GeV/$c^2$ ($J/ \psi$ rejection).
   
Different techniques are then used to identify the two $B$ mesons in an event. 
For the first ($B_{\rec}$) we select $\Bz \rightarrow \dstarelnubulo$ events 
with partial reconstruction of the decay $\Dstarm \rightarrow \Dzb \pi^-_s$,
using only the charged lepton from the \Bz\ decay and the soft pion
($\pi^-_s$) from the \Dstarm decay. The \Dzb\ decay is not reconstructed, 
resulting in high selection efficiency.
Electrons are identified using $dE/dx$ measurements in the DCH, the 
DIRC information and the ratio between the energy deposited in the EMC
crystals and the measured momentum in the DCH. Muons are identified using
the $dE/dx$ measurements in the DCH, the DIRC information and the number of 
hits in the IFR. 
Due to the limited phase space available in the \Dstar decay,
the \psoft is emitted within a cone with a half-opening-angle of 
approximately one radian and centered about the \Dstar motion direction
\cite{myframe}.
We approximate the direction of the \Dstar to be that of the \psoft
and estimate the energy  $\tilde{E}_{D^{*}}$ of the \Dstar
as a linear function of the energy of the \psoft, with parameters taken from 
the simulation. We define the square of the missing neutrino mass as:
\begin{equation}
\label{eq:m2}
\mnu = \left( \frac{\sqrt{s}}{2} - \tilde{E}_{\Dstar} - E_{\ell} \right)^2 -
(\tilde{\bf{p}}_{\Dstar} + {\bf{p}}_{\ell} )^2 ,
\end{equation}
where all quantities are defined in the \FourS\ frame. We neglect 
the momentum of the \Bz\ (approximately 0.34 GeV/$c$),
and identify the \Bz\ energy with half the total energy of the events 
($\sqrt{s}/2$).
$E_{\ell}$ and ${\bf{p}}_{\ell}$ are the energy and momentum vector 
of the lepton and
$\tilde{\bf{p}}_{\Dstar}$ is the estimated momentum vector of the \Dstar.
The distribution of $\mnu$ is peaked for signal events, while it is 
spread over a wide range for background events.

We determine the \Bz\ decay point from a vertex fit of the $\ell$ and \psoft
tracks, constrained to the beam-spot position in the plane perpendicular 
to the beam axis (the $x$-$y$ plane). The beam spot position and size 
are determined on a run-by-run basis using two-prong events
\cite{ref:babar}. Its size in the horizontal ($x$) direction is on average
120 $\mu$m. Although the beam spot size in the vertical ($y$) direction is only
5.6 $\mu$m, we use a constraint of 50 $\mu$m in the vertex fit
to account for the flight distance of the \Bz\ in the $x$-$y$ plane.

To suppress leptons from charm meson decays, we use only 
high-momentum leptons in the range $ 1.3 < p_{\ell} < 2.4 $ GeV/$c$.
The \psoft candidates have momenta ($p_{\pi_s}$)
between 60 and 200 \mevc. We 
reject events for which the $\chi^2$ probability of the vertex fit, 
${\cal P}_V$, is less than $0.1\%$.
We do not use dedicated optimization procedures for these cuts,
rather we apply a selection criterion to a  
likelihood ratio, ${\cal X}$, calculated from the signal and background 
distributions of $p_{\ell}$, $p_{\pi_s}$, and
${\cal P}_V$. We reject events for which ${\cal X}$ is lower than~$0.3$ and
we retain the $\ell-\pi_s$ pair with the highest value of ${\cal X}$ when
more than one candidate is found.

The flavor of the other $B$ meson ($B_{\tag}$) is determined by 
means of lepton tagging ($\ell = e$, $\mu$).  
A cut on the momentum in the center-of-mass frame is applied:
$1.0 < p_{\ell} < 2.35$ ($1.1 < p_{\ell} < 2.35$) 
GeV/$c$ for electrons (muons).
If more than one lepton track in an event survives, the one with
the largest momentum is used to determine the flavor of the $B_\tag$.
The  track selected for tagging is used to compute the tag vertex
position, using a procedure analogous to that described above.
We finally require $|\Delta z| < 3$ mm and $0 < \sigma_{\Delta z} < 0.5$ mm.

\begin{figure}[b!thp]
\begin{center}
\includegraphics[width=8cm]{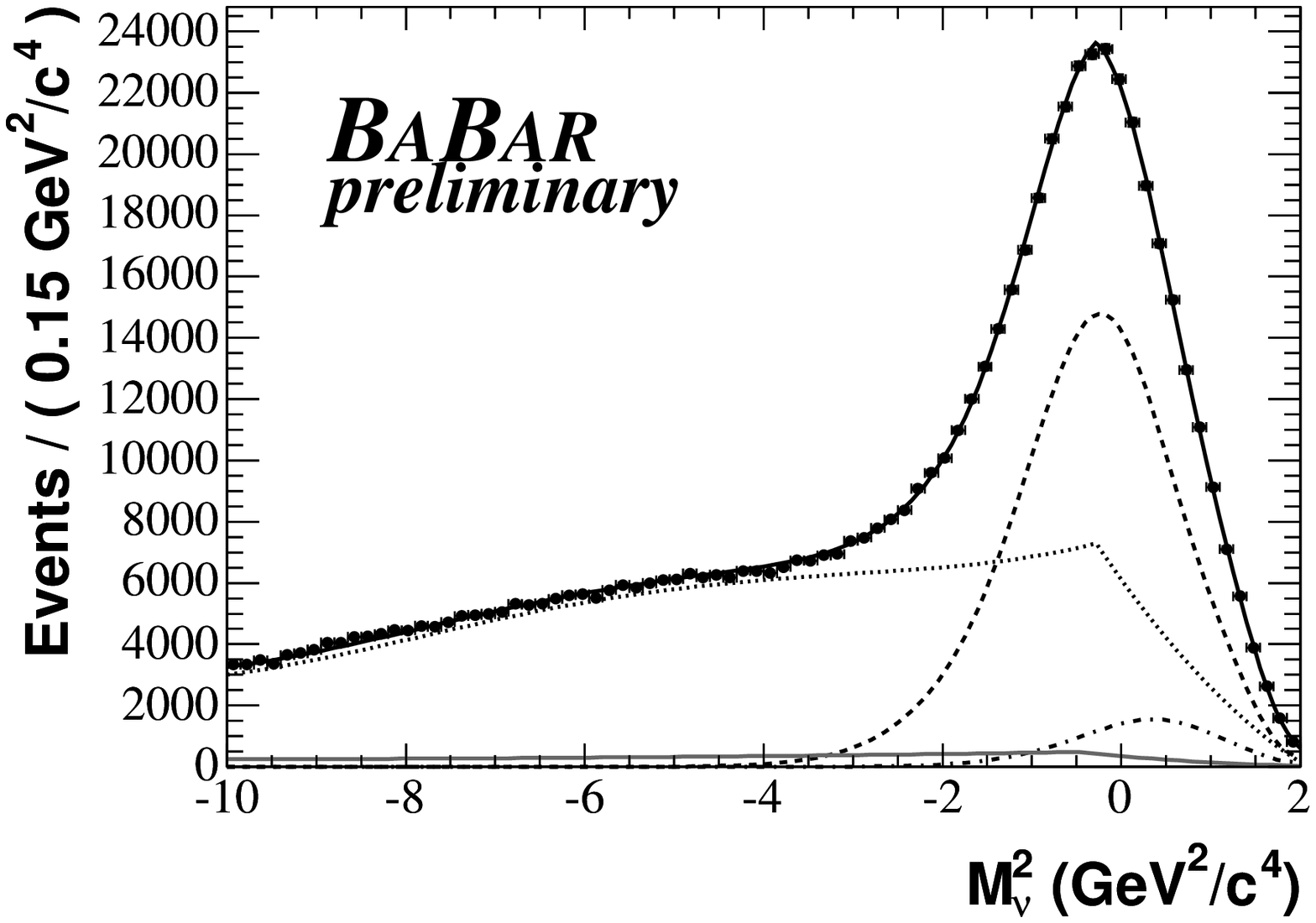} \\
\includegraphics[width=8cm]{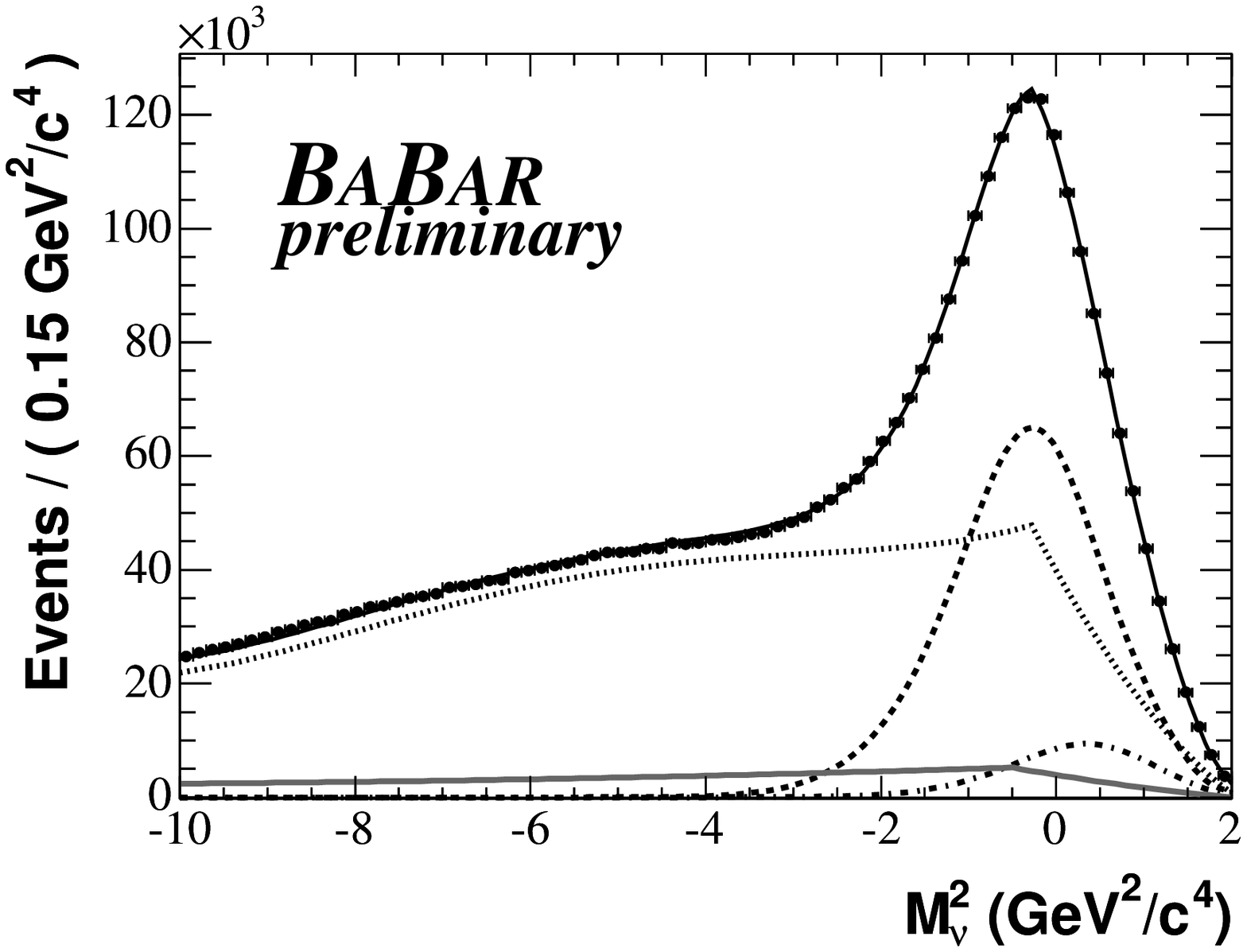} \\
\includegraphics[width=8cm]{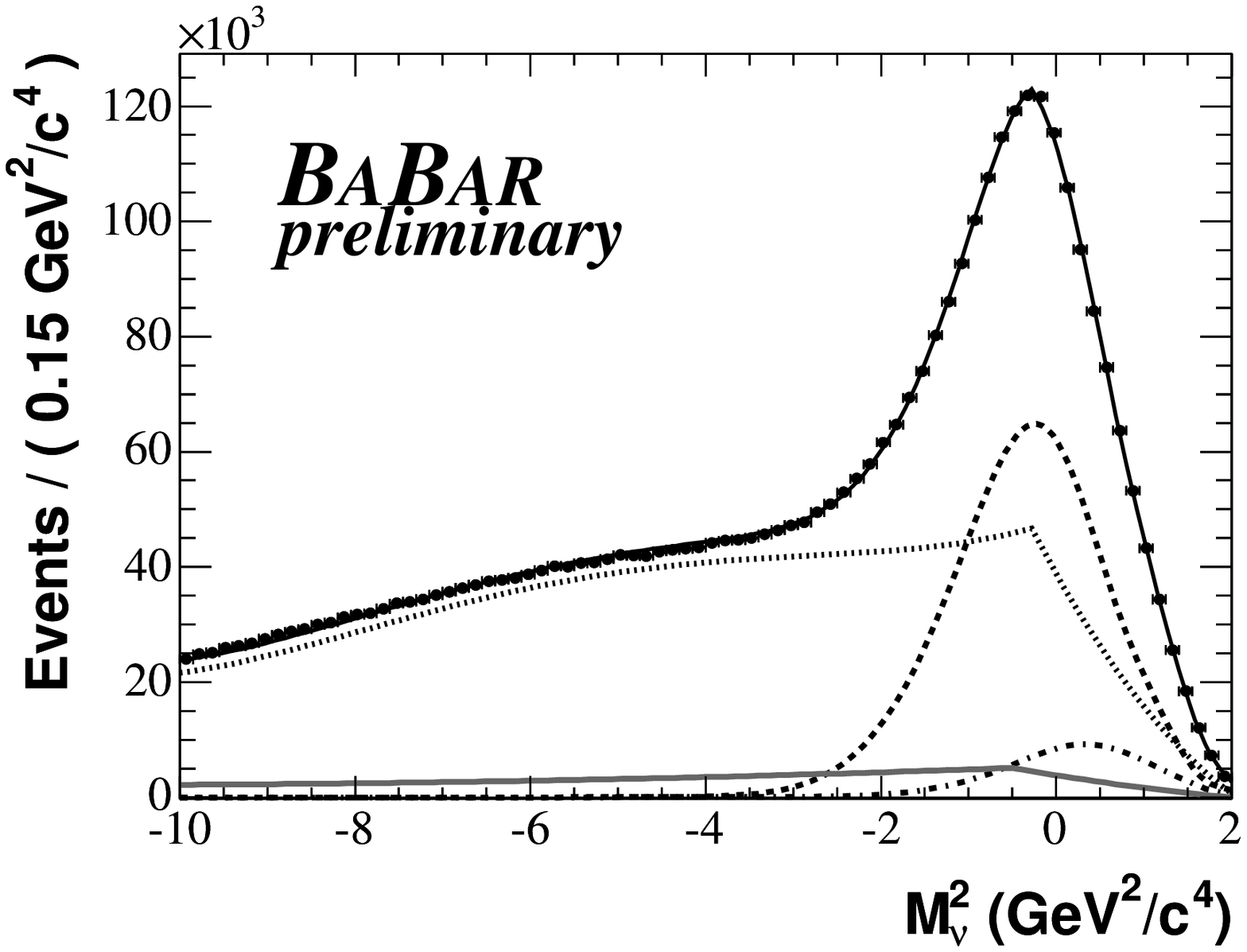}
\end{center}
\caption{Squared neutrino mass distributions for the tagged (top), \Bz\ 
untagged (center) and \Bzb\ untagged (bottom) data events. The following 
fitted contributions are superimposed: continuum (solid grey line), 
combinatorial (dotted), $B^{\pm}$ peaking (dash-dotted), signal (dashed),
all (solid black).}
\label{fig:mnucomp}
\end{figure}

These selection criteria accept 470,877 events in the data 
sample, which we will refer to as {\it tagged} events,
and 5,291,868 partially reconstructed events that fail only the
requirements of lepton tagging, which we will refer to as {\it untagged} 
events, and to \Bz\ untagged if the $B_\rec$ is reconstructed as
a \Bz\ and \Bzb\ untagged if the $B_\rec$ is reconstructed as a \Bzb. 
Fig.~\ref{fig:mnucomp} shows
the distributions of the squared neutrino mass for tagged, \Bz\ untagged
and \Bzb\ untagged events. 
Different components are recognized in the total sample:
{\it signal} events, due to \dstarelnubulo\ events, including the 
radiative decays \dstarelnubulo $\gamma / \pi^0 $ 
and other decays with the same final  
signature, like $B^0\to D^\ast \tau/X_c ~ (\tau/X_c 
\to \ell X)$; {\it peaking} background events, that correspond to decays of 
charged $B$ mesons which peak in the \mnu\ distribution, 
as the signal, like $B^+ \to D^{\ast\ast 0} \ell^+ \nu_{\ell}$, $
D^{\ast\ast 0} \to D^{*-} \pi^+$;
{\it combinatorial} events, corresponding to all $\BB$ decays not included 
in the previous categories; {\it continuum} events, coming from light quark
decays $e^+e^- \to q\bar{q}$ with $ q=u,d,c,s$ or $e^+e^- \to \ell^+ \ell^-$  
with $\ell=e,\mu,\tau$. 

The following procedure is used to determine the fractions of the 
various components in data. The \mnu\ distributions are fitted separately for
the different types of events, using Monte Carlo
events for the \BB\ components and off-resonance events for the continuum
component. The probability density functions (PDFs) for the different 
components are a Gaussian plus a function $\mathcal{B}(\mnu)$
\cite{myfunction}
for signal and peaking background, and piecewise continuous polynomials for 
combinatorial and continuum samples.
For signal events, we find the distribution peaking slightly below zero
mainly because of soft pion decays in flight and 
electron energy losses; as for the peaking 
background, the shift to larger positive values is due to the fact that in 
this case we neglect the presence of one (or more) additional particles.
We then fit the data sample, fixing all the shape parameters for
peaking, combinatorial and continuum background. The fraction
of continuum $f^\mathrm{cont}$ is also fixed, using the ratio between 
on-resonance and off-resonance luminosities, as well as the fraction of
peaking background $f^\mathrm{peak}$,
that is extracted from the fit on Monte Carlo. Conversely we 
float the signal shape parameters and the fraction of signal $f^\mathrm{sig}$, 
determining also the fraction of combinatorial background 
$f^{\mathrm{comb}}= 1-f^{\mathrm{sig}}-f^{\mathrm{peak}} - f^\mathrm{cont}$. 

The fractions determined with this procedure are summarized in 
Table~\ref{tab:fractions} for the different data samples.
The \mnu\ region $-4.0 < \mnu < 2.0~\mathrm{GeV}^2/c^4$ is defined 
as {\it signal region} and will be used in the nominal fits from now on. The
remaining part ($-10.0 < \mnu < -4.0~\mathrm{GeV}^2/c^4$) will be referred to
as {\it side-band region} and will be used in some fits to $\Delta t$ 
to determine combinatorial background parameters. Table~\ref{tab:fractions}
also shows fractions extrapolated to the signal region only. 
\begin{table}[h!tbp]
\caption{Fractions of signal, continuum, peaking and combinatorial background,
extracted from \mnu\ fits for the tagged, \Bz\ untagged and \Bzb\ untagged 
samples. The first 
uncertainties refer to the statistical uncertainty from the fit, 
the second are systematic errors from the shape parameter fixing.}
\begin{center}
\begin{tabular}{llcc}
\hline
Sample & Parameter & Fraction in the & Fraction extrapolated \\
       &   &  whole \mnu\ region & to the signal region \\
\hline
Tagged & $f^{\mathrm{sig}}$ & (32.72 $\pm$ 0.15 $\pm$ 0.58)\% & (46.59 $\pm$ 0.21 $\pm$ 0.82)\% \\ 
 & $f^{\mathrm{peak}}$ &  4.03\% & 5.54\%  \\
 & $f^{\mathrm{cont}}$ & 4.12\%  & 3.14\% \\
 & $f^{\mathrm{comb}}$ & (59.13 $\pm$ 0.15 $\pm$ 0.58)\% & (44.73 $\pm$ 0.21 $\pm$ 0.82)\% \\
\hline
Untagged $+$ & $f^{\mathrm{sig}}$ & (23.51 $\pm$ 0.03 $\pm$ 0.27)\% & (35.72 $\pm$ 0.05 $\pm$ 0.41)\% \\
 & $f^{\mathrm{peak}}$ & 3.05\% & 4.47\%   \\
 & $f^{\mathrm{cont}}$ & 6.47\%  & 5.38\% \\
 & $f^{\mathrm{comb}}$ & (66.97 $\pm$ 0.03 $\pm$ 0.27)\% & (54.43 $\pm$ 0.05 $\pm$ 0.41)\% \\
\hline
Untagged $-$ & $f^{\mathrm{sig}}$ & (24.23 $\pm$ 0.04 $\pm$ 0.20)\% & (36.33 $\pm$ 0.06 $\pm$ 0.30)\% \\
 & $f^{\mathrm{peak}}$ & 3.02\% & 4.42\% \\
 & $f^{\mathrm{cont}}$ & 6.14\%  & 5.20\%  \\
 & $f^{\mathrm{comb}}$ & (66.61 $\pm$ 0.04 $\pm$ 0.20)\% & (54.05 $\pm$ 0.06 $\pm$ 0.30)\% \\
\hline
\end{tabular}
\end{center}
\label{tab:fractions}
\end{table}
 
\section{TIME-DEPENDENT ANALYSIS}
\label{sec:TDanalysis}

Since the \Bz\ mesons can undergo oscillation, we will refer to {\it mixed}
events when both decay in the same flavor state ($B^0B^0$ or
$\kern 0.18em\overline{\kern -0.18em B}{}^0 
\kern 0.18em\overline{\kern -0.18em B}{}^0$),
and define the mixing state $s_m = -1$ for such events. Conversely, 
we will refer to {\it unmixed} events ($s_m = 1$) if the flavors of the two
$B$ mesons are opposite. Similarly we define $s_t = 1$ if the $B_{\tag}$ 
is a \Bz, $s_t = -1$ if the $B_{\tag}$ is a \Bzb.

The calculation of the decay rate for the four
tag-mixing states is straightforward, since we are dealing with semileptonic
\Bz\ decays both on the reconstructed and tag side and these are pure flavor
eigenstates:
\begin{eqnarray}
\nonumber
 && s_m= -1:~~~~~~~\mathcal{F}(\Delta t) = \frac{1}{2\tau_{B^0}} 
e^{-\frac{|\Delta t|}{\tau_{B^0}}} \bigg\vert\frac{q}{p}\bigg\vert^{-2s_t} 
[1 - \cos(\deltamd \Delta t)] \\
&& s_m= 1:~~~~~~~~~~~
\mathcal{F}(\Delta t) =  \frac{1}{2\tau_{B^0}} e^{-\frac{|\Delta t|}{\tau_{B^0}}}
[1 + \cos(\deltamd \Delta t)]
\label{eq:teoric}
\end{eqnarray} 
where \deltamd\ is the \Bz\ oscillation frequency, $\tau_{B^0}$ is the average 
lifetime of the physical states and we set $\Delta\Gamma$, 
the lifetime difference between the physical states, equal to 0. \\

For the $\Delta t$ distribution of signal events we use 
Eq.~(\ref{eq:teoric}), modified to account for several experimental 
effects.

The $B$-tagging algorithm introduces a modification to the theoretical
PDFs as there is a finite probability per event (called the
{\it mistag rate}) of falsely tagging with a wrong-sign lepton candidate. 
Considering a 
\Bz\ mistag rate (i.e. the probability for a true $\Bz$ being 
tagged as a \Bzb) and a \Bzb\ mistag
rate (i.e. the probability for a true $\Bzb$ being tagged as a \Bz),
we define ${w}$ as the average value of these probabilities and 
$\Delta{w}$ as their difference. 
Also, the signal PDFs are modified to reflect the fact that reconstruction
efficiency for $\ell^+ \pi_s^-$ pairs can be different from the
reconstruction efficiency for $\ell^- \pi_s^+$ pairs.
We therefore define an average reconstruction efficiency $\varepsilon_{rec}$ 
and a {\it reconstruction asymmetry} $A_{rec}$.
In the same way, the tagging 
efficiency for positive leptons can be different from the
tagging efficiency for negative leptons, so
we define an average
tagging efficiency $\varepsilon_{tag}$ and a {\it tagging asymmetry} $A_{tag}$.
We normalize our distributions to an average reconstruction efficiency
$\varepsilon_{rec} = 1$.

To account for the finite resolution of vertex determinations,
the function  of the true lifetime difference
$\Delta t_{true}$ must be convolved with a decay time difference 
resolution function.
We adopt a 3-Gaussian description where the two main contributions 
(referred to as ``narrow'' and ``wide'') depend
on the per-event error of the vertex separation, $  \sigma_{\Delta t} $,
while the third (``outlier'' Gaussian) is independent of 
$  \sigma_{\Delta t} $:

\begin{eqnarray}
\nonumber 
&& \mathcal{R}(b_n,s_n,b_w,s_w,s_o,f_w,f_o) = f_o \frac{1}{\sqrt{2\pi}s_o}e^{-\frac{t^2}{2 s_o^2}}
 + \\
&& + f_w \frac{1}{\sqrt{2\pi}s_w \sigma_{\Delta t}}e^{-\frac{(t - b_w \sigma_{\Delta t})^2}{2(s_w \sigma_{\Delta t})^2}} 
+ (1-f_w-f_o)\frac{1}{\sqrt{2\pi}s_n \sigma_{\Delta t}}e^{-\frac{(t - b_n \sigma_{\Delta t})^2}{2(s_n \sigma_{\Delta t})^2}}
\end{eqnarray}
where $t \equiv \Delta t - \Delta t_{true}$, $b_i$ and $s_i$ ($i = n,w$) are 
respectively the biases and the error scale factors for the narrow and
wide components, while $s_o$ is the standard deviation of the outlier
Gaussian (for this component we assume $b_o = 0$); $f_w$ and $f_o$ are
respectively the fractions of wide and outlier Gaussians.
 
Accounting for these effects, we define a PDF for signal events and
{\it direct} leptons, meaning tagging leptons that
do not originate from secondary charmed meson decay:  
\begin{equation}
\mathcal{F}_{\mathrm{dir}}^{\mathrm{sig}}(\Delta t, \sigma_{\Delta t}|s_m,s_t)= 
\mathcal{F}^{meas} (\Delta t,\sigma_{\Delta t} |s_m,s_t) \otimes 
\mathcal{R}_{\mathrm{dir}}^{\mathrm{sig}}
\end{equation}
where, using the definition $k = |q/p| -1$:
\begin{eqnarray}
\nonumber
 s_m= -1:~~ && \mathcal{F}^{\mathrm{meas}}(\Delta t) = \frac{\varepsilon_{tag}}{4\tau_{B^0}}(1+s_t A_{rec}) e^{-\frac{|\Delta t|}{\tau_{B^0}}} \{[1-s_t(\Delta{w}+2k(1-{w})-A_{tag}(1-2{w}))] \\
\nonumber
 && - [1-2{w}+s_t(A_{tag} -2k(1-{w}))]\cos(\deltamd \Delta t)\} \\
\label{eq:signdir}
\end{eqnarray}
\begin{eqnarray}
\nonumber
s_m= 1:~~~~~~~ && \mathcal{F}^{\mathrm{meas}}(\Delta t) = \frac{\varepsilon_{tag}}{4\tau_{B^0}} (1-s_t A_{rec}) e^{-\frac{|\Delta t|}{\tau_{B^0}}} \{ [1-s_t(\Delta{w}-2k{w}-A_{tag}(1-2{w}))] \\
&& + [1-2{w}+s_t(A_{tag} + 2k{w})]\cos(\deltamd \Delta t)\}
\label{eq:signdir2}
\end{eqnarray} 

For signal events in which the tagging lepton comes from a secondary charm
decay ({\it cascade} leptons), we use different PDFs according to the two 
possibilities that the
charm meson is a product of the $B_\tag$ or is the unreconstructed \Dzb\ on 
the decay side ({\it decay-side tagged} events). In the former case we just 
take the signal PDFs with the
reversed tag information: $\mathcal{F}_{\mathrm{cas}}^{\mathrm{sig}}(\Delta t,
\sigma_{\Delta t}|s_m,s_t)= \mathcal{F}_{\mathrm{dir}}^{\mathrm{sig}}
(\Delta t,\sigma_{\Delta t} |-s_m,-s_t)$ and we allow for different mistag and resolution
parameters. In the latter case, we use a simple exponential term:    
\begin{equation}
\mathcal{F}_{\mathrm{Dtag}}(\Delta t, \sigma_{\Delta t}|s_m,s_t)= 
\frac{1}{2 \tau_{D_e}}e^{-\frac{|\Delta t|}{\tau_{D_e}}}(1+s_m D_{\mathrm{Dtag}})(1+s_t A'_{tag}) (1 - s_t s_m A_{rec}) \otimes \mathcal{R}_{\mathrm{Dtag}}
\label{eq:dtag}
\end{equation}
where $ \tau_{D_e}$ is an effective \Dz\ lifetime and $ A'_{tag}$ is a 
distinct tagging asymmetry for this sample (see Sec.~\ref{sec:simu}). We 
characterize from now
on the mistag in a given sample $j$ in terms of the {\it dilution} parameter 
$D_j = 1-2{w}_j$. 

For charged $B$ peaking background events, we similarly use a pure 
lifetime PDF for the direct component:
\begin{equation}
\mathcal{F}^{+}_{\mathrm{dir}}(\Delta t, \sigma_{\Delta t}|s_m,s_t)= 
\frac{1}{2 \tau_{B^+}}e^{-\frac{|\Delta t|}{\tau_{B^+}}}(1+s_m D_{+})(1+s_t A_{tag}) (1 - s_t s_m A_{rec}) \otimes \mathcal{R}_{\mathrm{dir}}^{\mathrm{sig}}
\label{eq:peaking}
\end{equation}
while we use Eq.~(\ref{eq:dtag}) for the decay-side tagged events.

For combinatorial background events, we recognize different components that can be 
parameterized with the PDFs defined above, but with sets of effective 
parameters that need to be determined from the \mnu\ sidebands. 
For the neutral part we take into account three contributions: first, the 
events in which both leptons in the events originate directly from the 
two $B$ mesons,
even if the reconstructed side is not a genuine \dstarelnubulo\ event. 
For example, these 
may be other semileptonic $B$ decays in which the reconstructed 
soft pion candidate
is actually a random track in the event. For these events we use 
Eqs.~(\ref{eq:signdir}) and (\ref{eq:signdir2}), where some shape 
parameters are allowed to differ from the corresponding
ones for signal, since the vertex position is mis-determined 
using a wrong pion track. We also allow $A_{rec}^{\mathrm{bkg}}$
to be distinct from $A_{rec}^{\mathrm{dir}}$ for the same reason, but we
use the same $CP$ asymmetry parameterization of the signal, since
this kind of events carries the correct tag-mixing information. 
Second, the same procedure applies to cascade combinatorial events.  
The third contribution comes from events in which the tag track is coming
directly from a $B$, while the lepton on the decay side comes from a 
secondary charm meson. For this kind of events we assume:
$\mathcal{F}_{\mathrm{cas,2}}^{\mathrm{bkg}}(\Delta t,
\sigma_{\Delta t}|s_m,s_t)= \mathcal{F}_{\mathrm{dir}}^{\mathrm{bkg}}
(\Delta t,\sigma_{\Delta t} |-s_m,s_t)$ and we allow for different mistag and resolution
parameters with respect to the direct lepton case.
For decay-side tagged and charged $B$ events the descriptions of 
Eqs.~(\ref{eq:dtag}) and (\ref{eq:peaking}) apply also for combinatorial
events, given the changes of parameterization for resolution and
asymmetries.

For continuum events we use a single lifetime distribution:
\begin{equation}
\mathcal{F}^{\mathrm{cont}}(\Delta t, \sigma_{\Delta t}|s_m,s_t)=
\frac{1}{2 \tau_{\mathrm{cont}}}e^{-\frac{|\Delta t|}{\tau_{\mathrm{cont}}}}(1+s_m D_{\mathrm{cont}})(1+s_t A^{\mathrm{cont}}_{tag}) (1 - s_t s_m A^{\mathrm{cont}}_{rec}) \otimes \mathcal{R}_{\mathrm{dir}}^{\mathrm{bkg}}
\label{eq:offpeak}
\end{equation}
Its parameters are all determined from a fit to off-resonance events. 

In summary, the $\Delta t$ total PDF will be a sum of all the terms so far
introduced:
\begin{eqnarray}
\nonumber
&& \mathcal{F}(\Delta t,\sigma_{\Delta t}|s_m,s_t) = \\
\nonumber
&& f^{\mathrm{sig}}\{(1-g^{\mathrm{sig}}_{\mathrm{Dtag}})[(1-
g^{\mathrm{sig}}_{\mathrm{cas}}) \mathcal{F}^{\mathrm{sig}}_{\mathrm{dir}}
+ g^{\mathrm{sig}}_{\mathrm{cas}} \mathcal{F}^{\mathrm{sig}}_{\mathrm{cas}}]
+ g^{\mathrm{sig}}_{\mathrm{Dtag}}\mathcal{F}_{\mathrm{Dtag}}\} \\
\nonumber
&& + f^{\mathrm{peak}} [(1-g^{\mathrm{sig}}_{\mathrm{Dtag}}) \mathcal{F}^{+}_{\mathrm{dir}}
+ g^{\mathrm{sig}}_{\mathrm{Dtag}} \mathcal{F}_{\mathrm{Dtag}}] \\
\nonumber
&& + f^{\mathrm{comb}}\{(1-g_+^{\mathrm{bkg}}-g^{\mathrm{bkg}}_{\mathrm{Dtag}})[(1-g^{\mathrm{bkg}}_{\mathrm{cas}}-g^{\mathrm{bkg}}_{\mathrm{cas,2}})
\mathcal{F}^{\mathrm{bkg}}_{\mathrm{dir}} \\
&&
\nonumber 
+ g^{\mathrm{bkg}}_{\mathrm{cas}}\mathcal{F}^{\mathrm{bkg}}_{\mathrm{cas}}
+ g^{\mathrm{bkg}}_{\mathrm{cas,2}}\mathcal{F}^{\mathrm{bkg}}_{\mathrm{cas,2}}] + 
g_+^{\mathrm{bkg}}\mathcal{F}^{\mathrm{bkg}}_{+} +
g^{\mathrm{bkg}}_{\mathrm{Dtag}}\mathcal{F}_{\mathrm{Dtag}}\} \\
&& + f^{\mathrm{cont}}\mathcal{F}^{\mathrm{cont}}  
\label{eq:total}
\end{eqnarray}
where $g_j^i$ are generic fractions of the tagging category $j$ in sample
$i$ and we assume the fractions of decay-side tagged events, determined from
Monte Carlo simulation, to be equal in the
signal and peaking background components.

A crucial point of this analysis is to determine particle detection
asymmetries, that could in principle be degenerate with the $CP$ asymmetry.
A method commonly used for determining detection asymmetries in time-dependent
analyses is to use tagged and untagged events reconstructed as \Bz\ or \Bzb,
to have four equations from which 
$\varepsilon_{tag}$, $\varepsilon_{rec}$, $A_{rec}$ and $A_{tag}$ are
determined uniquely. This procedure, fully explained in Appendix of \cite{fmv},
is however appropriate when $CP$ violation in mixing is assumed to be zero.
Here we must employ a different approach. We parameterize the total number 
of expected events for each tagged
and untagged category in a likelihood function, which 
contains a term coming from the shapes of the PDFs for tagged events,
an extended term for the number of tagged events and an extended term for 
the number of untagged events. We show the procedure in full detail for 
signal events.     
The expected number of tagged events per category is obtained in a 
straightforward way by integrating Eqs.~(\ref{eq:signdir}) and 
(\ref{eq:signdir2}), while for untagged events:
\begin{equation}
N_{exp}(s_u) = N_{tot}\frac{1-s_u A_{rec}}{2} [1-\varepsilon_{tag}+s_u \varepsilon_{tag} A_{tag} x_d -s_u k(1-x_d)(1-\varepsilon_{tag})]
\end{equation}
where $N_{tot}$ is the total number of events (tagged and untagged),
$x_d = 1/[1+(\tau_{B^0}\Delta m_d)^2]$ and $s_u = +1 (-1)$ for \Bz\ (\Bzb) 
untagged events.

The likelihood function to be maximized is of the form:
\begin{equation}
\ln L_{tot} = \ln L - \sum_{s_t,s_m} \ln L^{ext}(s_t,s_m) - \sum_{s_u} \ln L^{ext}(s_u),
\end{equation}
where $\ln L^{ext}(i,(j)) = \ln(N_{obs}^{i,(j)}!) - N_{obs}^{i,(j)} 
\ln N_{exp}^{i,(j)} + N_{exp}^{i,(j)}$
are the extended terms for tagged and untagged events while $\ln L$ is the 
log-likelihood constructed starting from Eq.~(\ref{eq:total}). This method 
reduces the correlation between $CP$
mixing asymmetry and particle detection asymmetry, that nevertheless remains
significant, as quoted in Sec.~\ref{sec:datafit}. 

\section{FITS OF SIMULATED SAMPLES}
\label{sec:simu}

Several tests were performed on simulated events to verify that the fit 
determines the free parameters $k$, $A_{rec}$ and 
$A_{tag}$ correctly. First, fits to the
various components of the Monte Carlo sample and to the total \BB\ sample 
were performed to check for the presence of biases in the analysis 
technique. This Monte Carlo sample was generated with a value of $k$ equal to
0. We summarize in Table~\ref{tab:fitres} the
result of these fits for side-band region (first column) and signal
region (second column). $\tau_{B^0}$, $\tau_{B^+}$ and \deltamd\ are
fixed to the generated values in this procedure, while the 
normalizations of the various contributions are fixed to the 
values extracted from the \mnu\ distribution. 

The most notable parameters in Table~\ref{tab:fitres} are the asymmetries 
$k$, $A_{rec}$, $A_{tag}$, $A_{rec}^{\mathrm{bkg}}$ and $A'_{tag}$. We
find all to be compatible with 0, except $A_{rec}^{\mathrm{bkg}}$ and 
$A'_{tag}$. The source of $ A_{rec}^{\mathrm{bkg}} \neq 0 $ is random tracks
reconstructed as soft pions in the combinatorial background. We do not use a
control sample for estimating $A_{rec}^{\mathrm{bkg}}$  but we use the value 
found in the side-band fit; its possible variation from side-band to
signal region is taken into account during systematic evaluation.
From Monte Carlo studies, we find that the source of
$A'_{tag} \neq 0 $ is
charged kaons from the unreconstructed
\Dzb\ faking muons. We compare the asymmetry value 
found in the Monte Carlo sample (second column of Table~\ref{tab:fitres})
with the value calculated in a charged kaon control sample
($\Dstar\ \rightarrow\ \Dz \pi$, $\Dz \rightarrow K \pi$ decays) from the 
integrated $K^+$ and $K^-$ efficiencies over the tag lepton momentum spectrum. 
This is found to be compatible with the fitted value ($A'_{tag,CS} = 
(18.0 \pm 0.5)\%$). We then conclude that the asymmetry can be considered as 
entirely due to this source and we fix the value of $A'_{tag}$ to the one 
found in the control sample. A similar procedure is then exploited in the data 
fit.

Other checks are performed using large numbers of toy experiments, 
each generated with statistics
corresponding to the size of the data sample.
A set of 120 toy experiments
is used to test the validity of the uncertainty calculated by the minimization
algorithm and to check that the final value of the negative-log-likelihood
is compatible with the value found in the data. Other Monte Carlo samples
are generated with values of $k$ or $A_{rec}$ or $A_{tag}$ equal to $\pm 0.01$
to check the feasibility of disentangling the asymmetries from one another.
The fits show that this disentanglement is possible.
Also, a $CP$ asymmetry different from 0 is generated in the generic 
\BzBzb\ sample to verify the sensitivity of the fit to values of $k \neq 0$.
We find $k = (8.9 \pm 2.9) \damt$ for a generated value of $10 \damt$ and $k = 
(-10.1 \pm 2.9) \damt$ for a generated value of $-10 \damt$.

\begin{table}
\caption{Top: results of the fits to the \BB\ Monte Carlo and
for: side-band region (first column), signal region (second column).
Results of the fits to the on-resonance sample for side-band
region (third column), signal region (fourth column). We also report 
asymmetrical errors for some fundamental parameters. A horizontal line
is used to separate signal from background parameters.
Bottom: Continuum parameters determined from the total off-resonance sample:
they are fixed in the nominal on-peak data fit.
}
\footnotesize
\begin{center}
\begin{tabular}{lcccc}
\hline
Parameter & Side-band & Signal region & Side-band & Signal region \\
           & fit (MC) & fit (MC) & fit (data) & fit (data) \\
\hline
$k$ & (0.8 $\pm$ 3.0)\damt & (-1.6 $\pm$ 1.7)\damt & (-2.1 $\pm$ 4.9)\damt
 & (6.0 $\pm$ 3.4)\damt [+3.4, -3.5]\\
$A_{rec}$ & - & (0.02 $\pm$ 0.06)\damt & - & (1.0 $\pm$ 1.5)\damt [+1.5, -1.5] \\
$A_{tag}$ & (-0.2 $\pm$ 2.4)\damt & (0.3 $\pm$ 1.1)\damt & (-0.2 $\pm$ 2.9)\damt & (3.1 $\pm$ 2.1)\damt [+2.1, -2.1] \\
$b^{\mathrm{dir}}$ & - & -0.049 $\pm$ 0.002  & - & -0.044 $\pm$ 0.008 \\
$s_n^{\mathrm{dir}}$ & - & 0.859 $\pm$ 0.007 & - & 1.15 $\pm$ 0.02 \\
$f_w$ & - & (9.6 $\pm$ 0.3)\%  & - & (9.2 $\pm$ 0.7)\% \\
$s_w$ & - & 1.90 $\pm$ 0.03  & - & 1.90 (fixed) \\
$s_o$ & - & (7.2 $\pm$ 0.5) ps  & - &  7.2 ps (fixed) \\
$f_o$ & - & (2.7 $\pm$ 0.3)\damt  & - & 2.7\damt (fixed) \\
${w}_{\mathrm{dir}}$ & - & (1.32 $\pm$ 0.04)\%  & - & 1.32\% (fixed) \\
$\Delta{w}_{\mathrm{dir}}$ & - & (-3.2 $\pm$ 7.8)\damq  & - & -3.2\damq (fixed) \\
$\tau_{D_e}$ & (0.226 $\pm$ 0.009) ps & (0.267 $\pm$ 0.008) ps & (0.342 $\pm$ 0.015) ps & (0.321 $\pm$ 0.010) ps \\ 
${w}_{\mathrm{cas}}$ & - & (20.8 $\pm$ 2.4)\% & - & 20.8\% (fixed) \\
$\Delta{w}_{\mathrm{cas}}$ & - & (1.0 $\pm$ 0.4)\% & - & 1.0\% (fixed) \\
$b_n^{\mathrm{cas}}$ & - & -0.33 $\pm$ 0.04 & - & -0.28 $\pm$ 0.09 \\
$b_w^{\mathrm{cas}}$ & - & -2.4 $\pm$ 0.3 & - & -2.4 (fixed) \\
$A'_{tag}$ & - & (18.9 $\pm$ 1.0)\% & - & 7.6\% (fixed, see text) \\
$D_{\mathrm{Dtag}}$ & - & (83.7 $\pm$ 2.0)\% & - & 83.7 (fixed) \\
$g^{\mathrm{sig}}_{\mathrm{cas}}$ & - & (6.67 $\pm$ 0.20)\%  & - & (5.0 $\pm$ 0.5)\% \\
$g^{\mathrm{sig}}_{\mathrm{Dtag}}$ & - & 5.62\% (fixed) & - & 5.62\% (fixed) \\
\hline 
$D_{+}$ & (80.6 $\pm$ 0.8)\%  & (90.1 $\pm$ 2.5)\% & (83.2 $\pm$ 1.8)\% & (83.4 $\pm$ 4.2)\% \\
$\tau_{B^0}^{\mathrm{bkg}}$ & (1.469 $\pm$ 0.017) ps & (1.28 $\pm$ 0.04) ps & (1.51 $\pm$ 0.03) ps & (1.53 $\pm$ 0.02) ps \\
$\deltamd^{\mathrm{bkg}}$ & (0.479 $\pm$ 0.007) $\mathrm{ps}^{-1}$ & (0.37 $\pm$ 0.04) $\mathrm{ps}^{-1}$ & (0.506 $\pm$ 0.010) $\mathrm{ps}^{-1}$ & (0.586 $\pm$ 0.010) $\mathrm{ps}^{-1}$ \\
$A_{rec}^{\mathrm{bkg}}$ & (3.9 $\pm$ 2.3)\damt & 3.9\damt (fixed) & (9.7 $\pm$ 2.9)\damt & 9.7\damt (fixed) \\
${w}_{\mathrm{dir}}^{\mathrm{bkg}}$ & (1.19 $\pm$ 0.07)\% & 1.19\% (fixed) & 1.19\% (fixed) & 1.19\% (fixed) \\
$\Delta{w}_{\mathrm{dir}}^{\mathrm{bkg}}$ & (1.4 $\pm$ 0.9)\damt & 1.4\damt (fixed) & 1.4\damt (fixed) & 1.4\damt (fixed) \\
$b^{\mathrm{dir}}_{\mathrm{bkg}}$ & 0.071 $\pm$ 0.036 &  -0.034 $\pm$ 0.030 & 0.013 $\pm$ 0.006 & -0.056 $\pm$ 0.008 \\
$s^{\mathrm{dir}}_{n,\mathrm{bkg}}$ & 0.837 $\pm$ 0.006 & 0.837 (fixed) & 0.850 $\pm$ 0.011 & 0.850 (fixed) \\
$f_w^{\mathrm{bkg}}$ & (7.6 $\pm$ 0.5)\% & (6.1 $\pm$ 0.5)\% & (12.7 $\pm$ 0.8)\% & (7.4 $\pm$ 1.5)\% \\
$x_d^{\mathrm{bkg}}$ & 0.205 $\pm$ 0.009 & 0.205 (fixed) & 0.420 $\pm$ 0.012 & 0.420 (fixed) \\
${w}_{\mathrm{cas}}^{\mathrm{bkg}}$ & (18.7 $\pm$ 0.6)\% & 18.7\% (fixed) & 18.7\% (fixed) & 18.7\% (fixed) \\
$\Delta{w}_{\mathrm{cas}}^{\mathrm{bkg}}$ & 0.017 $\pm$ 0.006 & 0.017 (fixed) & 0.017 (fixed) & 0.017 (fixed) \\
$b^{\mathrm{cas}}_{n,\mathrm{bkg}}$ & -0.02 $\pm$ 0.06 & -0.02 (fixed) & -0.13 $\pm$ 0.08 & -0.13 (fixed) \\
$b^{\mathrm{cas},2}_{n,\mathrm{bkg}}$ & 0.02 $\pm$ 0.08 & 0.02 (fixed) & 0.15 $\pm$ 0.11 & 0.15 (fixed) \\
$g^{\mathrm{bkg}}_{\mathrm{Dtag}}$ & (13.0 $\pm$ 0.2)\% & (9.2 $\pm$ 0.8)\% & (8.5 $\pm$ 0.5)\% & (6.1 $\pm$ 0.8)\% \\
$g^{\mathrm{bkg}}_{+}$ & (40.7 $\pm$ 0.3)\% & (37.1 $\pm$ 0.2)\% & (33.7 $\pm$ 0.5)\% & (31.9 $\pm$ 0.3)\% \\
\hline
\end{tabular}
\\[5mm]
\begin{tabular}{lc}
\hline
Parameter & Value from \\
           & off-peak fit \\
\hline
$\tau_{\mathrm{cont}}$ & (0.62 $\pm$ 0.03) ps \\ 
$A_{rec}^{\mathrm{cont}}$ & (6.4 $\pm$ 2.1)\%  \\
$A_{tag}^{\mathrm{cont}}$ & (3.9 $\pm$ 2.9)\% \\
$D_{\mathrm{cont}}$ & (47.8 $\pm$ 1.6)\% \\
\hline
\end{tabular}
\end{center}
\label{tab:fitres}
\end{table}  

\section{FITS OF THE DATA SAMPLE}
\label{sec:datafit}
The same procedure used in Monte Carlo events is applied to the 
on-resonance data. We float the same set of
parameters and we fix the others in the following way. 
The tagging efficiency $\varepsilon_{tag}$
is determined from the number of tagged events divided by the total number of
events: we find $\varepsilon_{tag} = (9.06 \pm 0.02)\%$. For continuum events
we use parameters taken from the off-resonance fit. 
For combinatorial background, some parameters are floated, some are fixed 
from a previous side-band fit (see Table~\ref{tab:fitres}). For signal and
peaking background, we take dilutions from Monte Carlo and we float most 
resolution parameters.
For $A'_{tag}$, we approximate its value using the estimate from the
control sample explained above, assuming the fraction of charged kaon 
mistagging is the 
same as in Monte Carlo (we tested this hypothesis, as described in Sec.~\ref{sec:Systematics}).
We obtain an asymmetry of $ A'_{tag} = (7.6 \pm
0.8)\%$ from the control sample.
$\tau_{B^0}$, \deltamd\ and $\tau_{B^+}$ are fixed to their 
PDG values \cite{PDG}; the former two are then floated to 
estimate the fit stability in data when these parameters are free.

\begin{figure}[b!thp]
\begin{center}
\includegraphics[width=11cm]{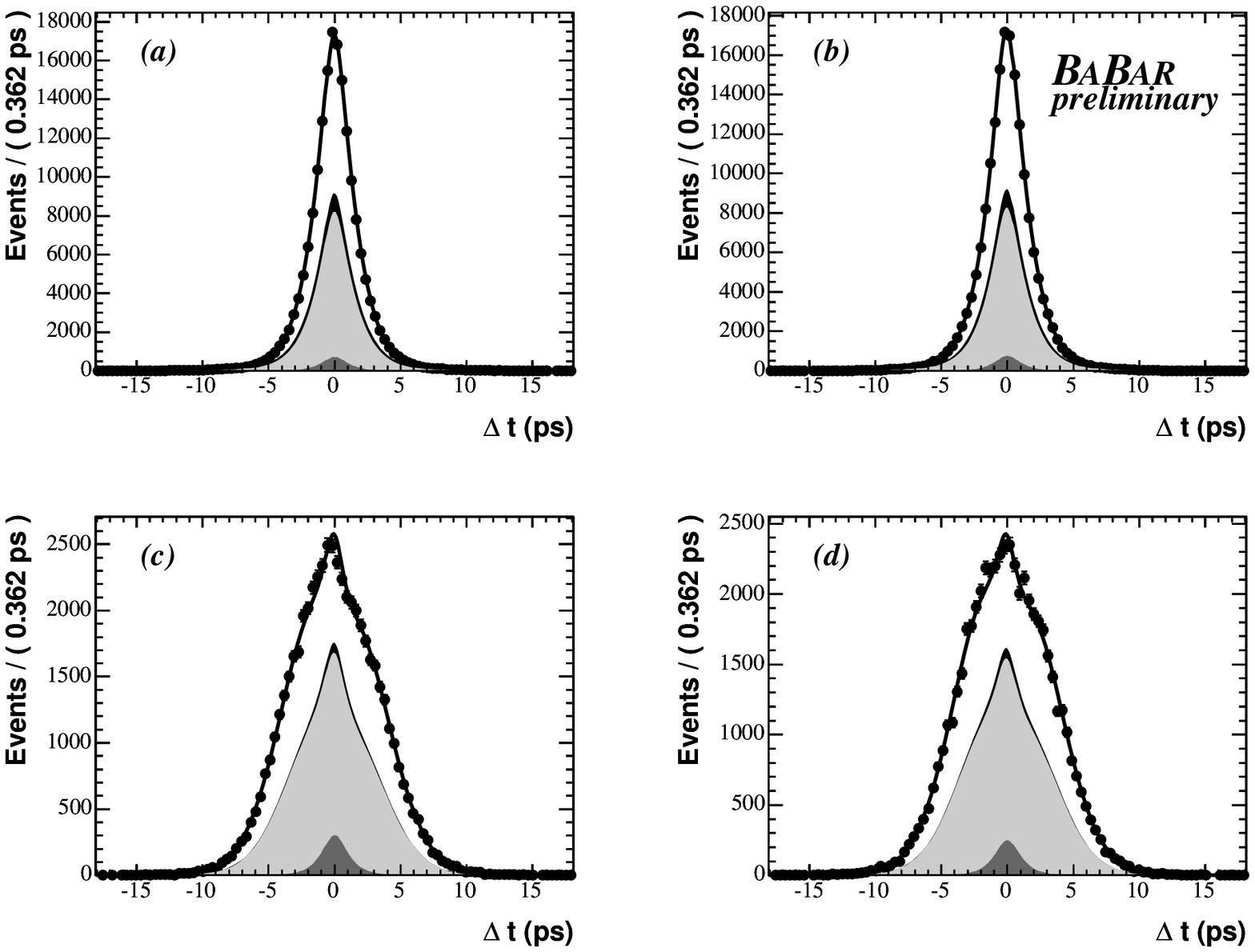} 
\end{center}
\caption{Fit to the $\Delta t$ distributions for the on-resonance data
samples in linear scale. Letters indicate
the different samples: (a) Unmixed positive $(s_m = 1, s_t = 1)$,
(b) Unmixed negative $(s_m = 1, s_t = -1)$,
(c) Mixed positive $(s_m = -1, s_t = 1)$,
(d) Mixed negative $(s_m = -1, s_t = -1)$. The following 
fitted contributions are shown in the fit: continuum (dark grey), 
combinatorial (light grey), $B^{\pm}$ peaking (black), signal (white).}
\label{fig:alldata}
\end{figure}

\begin{figure}[b!thp]
\begin{center}
\includegraphics[width=11cm]{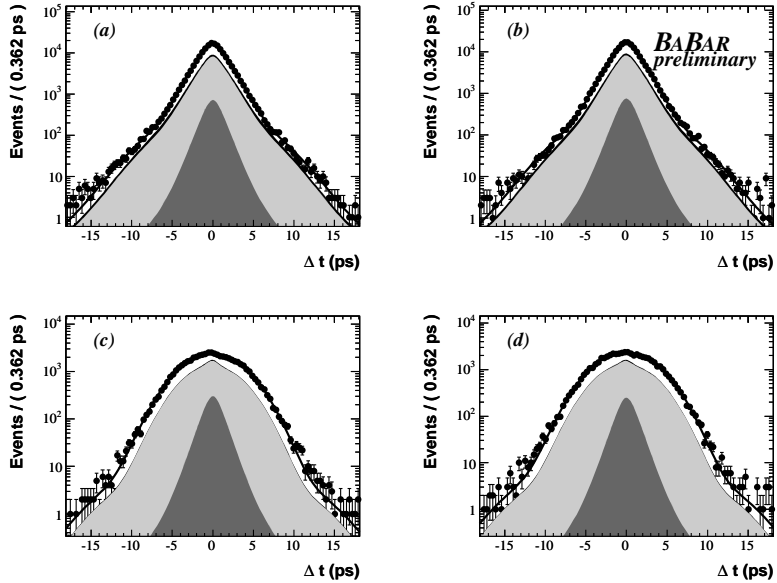} 
\end{center}
\caption{Same as Fig.~\ref{fig:alldata} in logarithmic scale.}
\label{fig:alldatalog}
\end{figure}

\begin{figure}[b!thp]
\begin{center}
\includegraphics[width=9cm]{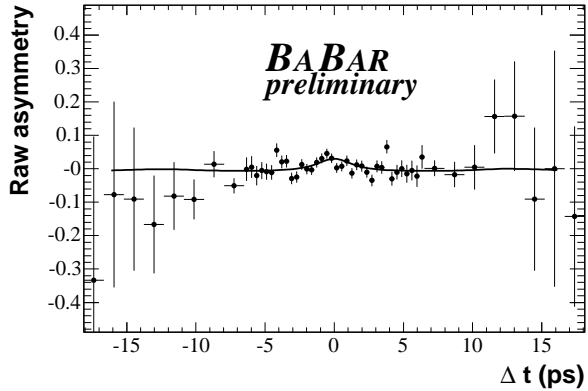}
\end{center}
\caption{Raw asymmetry between mixed positive $(s_m = -1, s_t = 1)$ and
mixed negative $(s_m = -1, s_t = -1)$ events with the PDF asymmetry
derived from the fit in Fig.~\ref{fig:alldata} superimposed.
The asymmetry at low values of $|\Delta t|$ corresponds to the kaon-mistag
contribution in decay-side tagged events (see text).}
\label{fig:asymdata}
\end{figure}

The fit results are shown in
Table~\ref{tab:fitres} (third and fourth column); for the three signal
asymmetry parameters we find the following correlation coefficients:
$\rho(A_{tag},A_{rec}) = 43.4\%$, $\rho(k,A_{rec}) = 50.2\%$, $\rho(k,A_{tag})
= 88.5\%$. Figs.~\ref{fig:alldata} and \ref{fig:alldatalog}
show the fitted PDFs for the subsamples equivalent to the four tag-mixing
states; Pearson's test gives a $\chi^2$ of 682 for 385 degrees of freedom. 
Fig.~\ref{fig:asymdata} shows the PDF asymmetry
derived from the fit in Fig.~\ref{fig:alldata} between
mixed events with $s_t = 1$ and $s_t = -1$.  
          
\section{SYSTEMATIC STUDIES}
\label{sec:Systematics}
We consider the following sources of systematic uncertainties. 

For the reconstruction asymmetry for combinatorial background events,
we repeat the fit in the first column of Table~\ref{tab:fitres} for 
combinatorial Monte Carlo events
in the signal region, finding $A_{rec}^{\mathrm{bkg}} = (6.5 \pm 2.1)\damt$.
The difference between this result and the one in the sideband
($= 2.6\damt$) is applied to the reconstruction asymmetry in data and
the fit is re-done to estimate the impact on $k$. 

We vary the continuum 
reconstruction and tagging asymmetries by $\pm 1\sigma$, where the
standard deviations are taken from the fit to the off-peak events.
The variations in $k$ are averaged and added in quadrature to give the 
systematic uncertainty.

We perform a fit in which the asymmetry $A'_{tag}$ is free rather than
fixed from a control sample: we obtain
$A'_{tag} = (4.0 \pm 3.8)\%$ which is compatible with the assumption
that the fraction of kaon mistagging is the same as in Monte Carlo, but
the uncertainties are large because of the high correlation between
$ A_{tag}$ and $ A'_{tag}$. The resulting variation in $k$ is taken as a 
systematic uncertainty.

We vary the sample fractions by $\pm 1\sigma$, where the statistical 
errors are taken from the \mnu\
data fits, both to tagged and untagged events separately. Moreover, 
for \Bz\ and \Bzb\ untagged events, we 
move the two central values in opposite directions in order to consider
the maximum asymmetry that could be generated by the fit uncertainty. 
Similarly, we vary the fractions of their correlated systematic uncertainties
simultaneously for tagged and untagged events.
Continuum fractions are varied by the uncertainty in the on/off-resonance 
luminosity ratio, which we take as 1.3\% from \cite{ref:babar}. 

For the analysis technique, we determine the mean of
$k_{\mathrm{fit}}-k_{\mathrm{gen}}$ in the Monte Carlo tests we performed and
we correct the result on data accordingly; the full size of the
correction ($5.5 \damq$) is taken as a systematic uncertainty.

We let $\tau_{B^0}$ and \deltamd\
float in the data fit, obtaining $\tau_{B^0} = (1.551 \pm 0.012)$ ps and
$ \deltamd = (0.475 \pm 0.019)~\mathrm{ps}^{-1}$ which are in good 
agreement with the world averages \cite{PDG}, considering a known bias
on \deltamd\ caused by the partial reconstruction technique \cite{margoni}.

Most mistag parameters 
for the various samples are fixed in the data fit. 
When omitting decay-side tagged
events, we can assume mistagging is almost entirely
due to charged pions. Using inclusive pion control samples, we notice 
that for both negative and positive pions, the relative difference between 
data and Monte Carlo is smaller than 10\%. We conservatively vary the mistags
of 10\% and quote the variations of $k$ as a systematic uncertainty.

Also some parameters of the resolution are fixed, namely the fraction and
width of the ``outlier'' Gaussian and the width of the ``wide'' Gaussian
are fixed. For the former, we repeat the fit setting $f_o = 0$ (2-Gaussian
resolution only), the latter is varied by 30\% conservatively.

All other parameters that are fixed from the sidebands or from the 
off-resonance fit are varied by their statistical errors to
estimate corresponding  systematic
uncertainties. 

All the variations in $k$ are summed in quadrature.            
The effect on the parameter $k$ for all these sources, as well as the total
uncertainty is summarized in Table \ref{tab:syst}.

\begin{table}[h!tbp]
\caption{Sources of systematic uncertainty for the $|q/p|$ measurement.}
\begin{center}
\begin{tabular}{lc}
\hline
Source & Syst. error ($\damt$)\\
\hline
Reconstruction asymmetry for combinatorial background & 1.1 \\
Asymmetries for continuum & 1.0 \\
Tagging asymmetry for decay-side tagged events & 0.2 \\
\mnu\ fractions                & 1.0 \\
Likelihood fit bias            & 0.6 \\
Physical parameter fixing & 0.5 \\
Mistag parameters        & 0.0 \\
Resolution function    & 0.3 \\
Sideband-fixed parameters & 0.5 \\
\hline
Total          & 2.0 \\
\hline
\end{tabular}
\end{center}
\label{tab:syst}
\end{table} 

\section{RESULTS}
\label{sec:Physics}
From the fit performed on the on-resonance data sample we measure:
\begin{equation}
\nonumber
|q/p| -1 = (6.5 \pm 3.4(\mathrm{stat.}) \pm 2.0(\mathrm{syst.})) \damt 
\end{equation}
which relates to $CP$ violation in mixing.

To facilitate comparison with other measurements
and formalisms, we also express our result
in two other commonly used notations.
The parameter $A_{SL}$ defined in Eq.~(\ref{eq:np})
is found to be:
\begin{equation}
\nonumber
A_{SL} = \frac{1-|q/p|^4}{1+|q/p|^4} = (-13.0 \pm 6.8(\mathrm{stat.}) \pm 4.0(\mathrm{syst.})) \damt 
\end{equation}
or, using the $\varepsilon_B$ parameter and the relation 
$q/p = (1-\varepsilon_B)/(1+\varepsilon_B)$:
\begin{equation}
\nonumber
\frac{\mathrm{Re}\varepsilon_B}{1+|\varepsilon_B|^2} \simeq \frac{A_{SL}}{4} = 
(-3.2 \pm 1.7(\mathrm{stat.}) \pm 1.0(\mathrm{syst.})) \damt
\end{equation}

\section{SUMMARY}
\label{sec:Summary}
We have presented a determination of the parameter $|q/p|$, that is a 
measurement of $CP$ violation in \BzBzb\ mixing, using a fit
to $\Delta t$, the time difference between the two $B$ decays. 
One of the $B$ mesons is partially reconstructed in the semileptonic 
channel \dstarelnubulo, i.e. only the lepton and the soft pion from
$\Dstarm \rightarrow \Dzb \pi^-$ decay are reconstructed, 
while the flavor of the 
other $B$ is determined by means of lepton tagging.    
We use a luminosity
of 200.8 $\mathrm{fb}^{-1}$ collected by the \babar\ detector in the period
1999-2004, and obtain the preliminary result:
\begin{equation}
\nonumber
|q/p| -1 = (6.5 \pm 3.4(\mathrm{stat.}) \pm 2.0(\mathrm{syst.})) \damt \, .
\end{equation}

This result is compatible with the current world average, and
the magnitude of its error
is comparable to those of the
most recent other measurements \cite{belle,yeche}. The 
corresponding central confidence interval at 95\% C.L. for $|q/p|$ is 
$[1.0012 - 1.0142]$. 

\section{ACKNOWLEDGMENTS}
\label{sec:Acknowledgments}

We are grateful for the 
extraordinary contributions of our \pep2\ colleagues in
achieving the excellent luminosity and machine conditions
that have made this work possible.
The success of this project also relies critically on the 
expertise and dedication of the computing organizations that 
support \babar.
The collaborating institutions wish to thank 
SLAC for its support and the kind hospitality extended to them. 
This work is supported by the
US Department of Energy
and National Science Foundation, the
Natural Sciences and Engineering Research Council (Canada),
Institute of High Energy Physics (China), the
Commissariat \`a l'Energie Atomique and
Institut National de Physique Nucl\'eaire et de Physique des Particules
(France), the
Bundesministerium f\"ur Bildung und Forschung and
Deutsche Forschungsgemeinschaft
(Germany), the
Istituto Nazionale di Fisica Nucleare (Italy),
the Foundation for Fundamental Research on Matter (The Netherlands),
the Research Council of Norway, the
Ministry of Science and Technology of the Russian Federation, and the
Particle Physics and Astronomy Research Council (United Kingdom). 
Individuals have received support from 
the Marie-Curie IEF program (European Union) and
the A. P. Sloan Foundation.


\begin{thebibliography}{99}

\bibitem{ciuch}
M.~Ciuchini {\it et al.}, JHEP {\bf 0308}, 031 (2002).

\bibitem{bene}
M.~Beneke {\it et al.}, Phys.\ Lett. {\bf B 576}, 173 (2003).

\bibitem{laplace}
S.~Laplace {\it et al.}, Phys.\ Rev. {\bf D}65, 094040 (2002).

\bibitem{ckm} 
M.~Kobayashi and T.~Maskawa, Prog.\ Theor.\ Phys.\  {\bf 49}, 652 (1973).

\bibitem{belle}
K.~Abe {\it et al.}, The BELLE Collaboration,
Phys.\ Rev. {\bf D}73, 112002 (2006).

\bibitem{yeche}
B.~Aubert {\it et al.}, The \babar\ Collaboration,
Phys.\ Rev.\ Lett. {\bf 96}, 251802 (2006).

\bibitem{fmv}
B.~Aubert {\it et al.}, The \babar\ Collaboration, Phys.\ Rev.\ {\bf D}70, 
012007 (2004).

\bibitem{dzero}
http://www-d0.fnal.gov/Run2Physics/WWW/results/prelim/B/B29/B29.pdf.

\bibitem{utfit}
M.~Bona {\it et al.}, The UT{\it fit} Collaboration, hep-ph/0605213.

\bibitem{ref:babar}
The \babar\ Collaboration, B.\ Aubert {\em et al.},
Nucl.\ Instrum.\ Methods {\bf A479}, 1-116 (2002).

\bibitem{foxwol}
G.~C.~Fox and S.~Wolfram, Phys.\ Rev.\ Lett. {\bf 41}, 1581 (1978).

\bibitem{myframe}
Unless specified, all four-momenta are measured in the \FourS\ rest frame.

\bibitem{myfunction}
With $m \equiv \mnu$, we define $\mathcal{B} (m) =
  [(m-m_0)^{a_1} + c(m-m_0)^{a_2}]e^{-b(m-m_0)}$. 

\bibitem{PDG}
Particle Data Group, 
S. Eidelman {\it et al.}, Phys.\ Lett.\  {\bf B592}, 1 (2004). 

\bibitem{margoni}
B.~Aubert {\it et al.}, The \babar\ Collaboration, Phys.\ Rev.\ {\bf D}73, 
012004 (2006).

\end{thebibliography}
\end{document}